\documentclass{article}
\usepackage{graphicx}
\usepackage{cite}
\usepackage{amssymb}
\usepackage[utf8]{inputenc}

\usepackage[T1]{fontenc}

\begin{document}
\title{RDM-stars and galactic rotation curves}
\author{Igor Nikitin\\
Fraunhofer Institute for Algorithms and Scientific Computing\\
Schloss Birlinghoven, 53757 Sankt Augustin, Germany\\
\\
igor.nikitin@scai.fraunhofer.de
}
\date{}
\maketitle

\begin{abstract}
The recently formulated model of black holes coupled to the radial flows of dark matter (RDM-stars) is considered and the shape of the galactic rotation curves predicted by the model is evaluated. Under the assumption that the density of black holes is proportional to the density of luminous matter, the model perfectly fits the experimental data, both for the universal rotation curve, describing the spiral galaxies of a general form, and for the grand rotation curve, describing the orbital velocities of Milky Way galaxy in a wide range of distances. The modeling of the galactic gravitational field at large distances from the center and out of the galactic plane is also discussed.
\end{abstract}

\section{Introduction}

The study of rotation curves of spiral galaxies, pioneered by Rubin et al. \cite{VR1970,VR1978,VR1980,VR1985}, has revealed the presence of hidden mass (dark matter) in the composition of galaxies. This topic been further investigated by Persic and Salucci \cite{9502091}, Persic et al. \cite{9503051,9506004}, Salucci et al. \cite{0703115}, Karukes and Salucci \cite{1609.06903}. In paper \cite{9506004} the concept of {\it universal rotation curve} (URC) has been put forward, describing the orbital velocities in the spiral galaxies of a general form by a single function of distance and luminosity. In further papers this concept has been extended to larger distances and a wider class of galaxies.

The rotation curves of spiral galaxies have also been studied by Sofue and Rubin \cite{0010594}. A special emphasis is done on the rotation curve of the Milky Way galaxy in papers by Sofue et al. \cite{0811.0859}, Sofue \cite{0811.0860,1110.4431,1307.8241}. In paper \cite{1307.8241} the rotation curve of the entire Milky Way galaxy has been constructed, called {\it grand rotation curve} (GRC). This curve covers the range of distances 1.1pc-1.6Mpc, from the central supermassive black hole to the intergalactic space. 

In the papers by Kirillov \cite{9911168,0203267,0211162,0505131}, Kirillov and Turaev \cite{0202302,0604496} a model of the universal coupling between dark and luminous matter has been developed, in a form of the integral (further referred as {\it KT-integral}):
\begin{eqnarray}
&&\rho_{dm}(x)=\int d^3x'\, b(x,x')\, \rho_{lm}(x'), \label{KT1}
\end{eqnarray}
where $\rho_{lm}$ is the mass density of luminous matter, $\rho_{dm}$ is the mass density of dark matter. Due to the symmetries, the kernel can be written in a form $b(x,x')=b(|x-x'|)$, with the following proposed dependence on the distance:
\begin{eqnarray}
&&b(r)=1/(4\pi L_{KT})/r^2, \label{KT2}
\end{eqnarray}
where $ L_ {KT} $ is a constant of the dimension of length. The physical interpretation of this formula is that each mass element of the luminous matter $ M_{lm} = (d ^ 3x \, \rho_{lm}) $ corresponds to a spherically symmetric distribution of dark matter with density and enclosed mass function of the form
\begin{eqnarray}
&&\rho_{dm}(r)=M_{lm}/(4\pi L_{KT})/r^2,\label{KT3}\\
&&M_{dm}(r)=4\pi \int_0^r dr' {r'}^2 \rho_{dm}(r')=M_{lm} r/L_{KT}.
\end{eqnarray}
That is, the enclosed dark matter mass depends on the distance linearly and $ L_ {KT} $ determines the slope of this dependence. Specifically, this is the distance at which the mass of dark matter equals to the mass of the luminous matter, to which it is coupled, $ M_ {dm} (L_ {KT}) = M_ {lm} $. The contribution of dark matter to the orbital acceleration, orbital velocity and gravitational potential is:
\begin{eqnarray}
&&a_{r,dm}=GM_{dm}(r)/r^2=GM_{lm}/(rL_{KT}),\\
&&v^2_{dm}=GM_{dm}(r)/r=GM_{lm}/L_{KT},\\
&&\varphi_{dm}=GM_{lm}/L_{KT}\cdot\log r. \label{phidm}
\end{eqnarray}
From here we see that for a point source or at large distances from a compact mass distribution, asymptotically flat rotation curves $v_{dm}=Const$ are reproduced.

\begin{figure}
\centering
\includegraphics[width=0.8\textwidth]{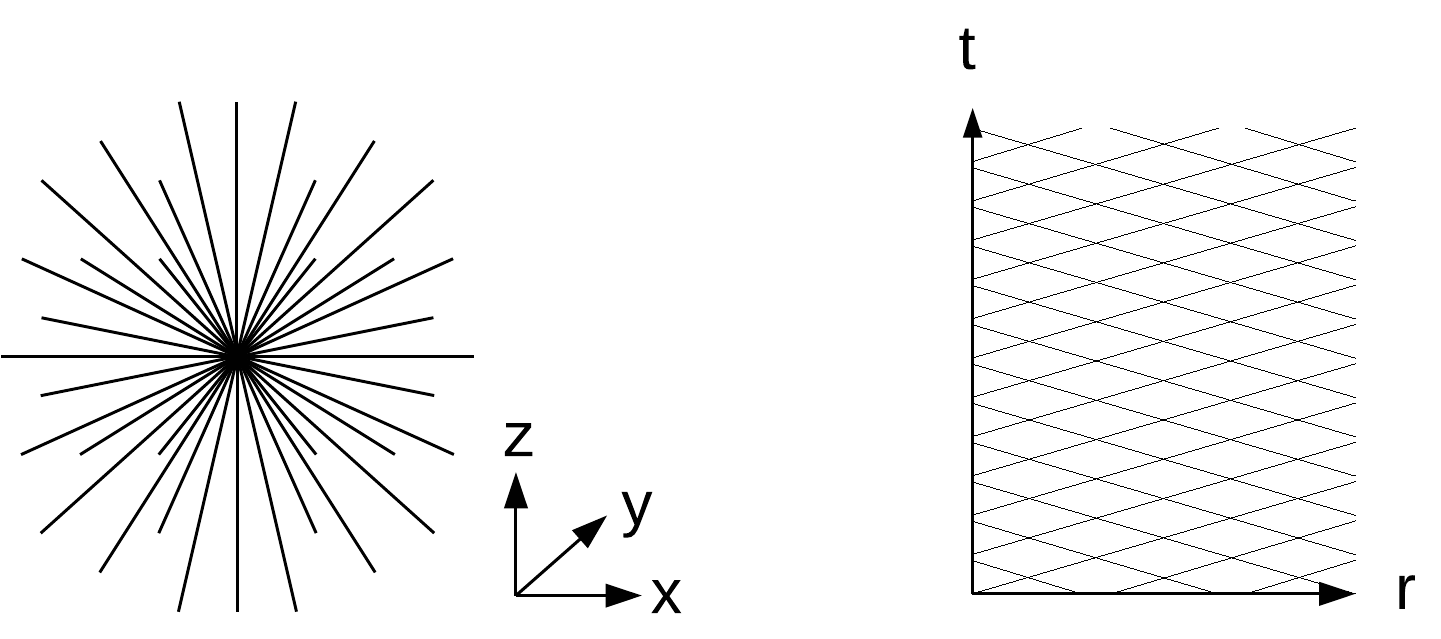}
\caption{RDM-star: a black hole, coupled to radial flows of dark matter, in $xyz$ and $tr$ coordinates. Image from~\cite{1701.01569}.}
\label{f0}
\end{figure}

In our papers \cite{1701.01569,1707.02764,1811.03368,1812.11801} a model of a black hole, coupled to radial flows of dark matter, {\it an RDM-star}, has been constructed. Due to its geometry with radially divergent flows, see Fig.\ref{f0}, the RDM-star possesses the dark matter density profile $\rho_{dm}\sim r^{-2}$ identical with the kernel of KT-integral. While a single RDM star produces the flat rotation curve, the distribution of RDM-stars in the galaxy modulates this dependence, generally producing the non-flat curve. If we assume that all black holes are RDM-stars and the density of black holes in the galaxy is proportional to the density of the luminous matter: $\rho_{bh}\sim\rho_{lm}$, then the distribution of the dark matter will be described by KT-integral (\ref{KT1}-\ref{KT2}) with all the resulting formulae (\ref{KT3}-\ref{phidm}). In this paper we will compute this integral for the available galactic models and compare the resulting rotation curves with the experimental data.

The idea behind the RDM-star geometry is that similarly to the stars emitting light, black holes emit dark matter. In more detail, since we consider a stationary T-symmetric model, black holes do not only emit, but also absorb dark matter in equal amounts, so that $tr$ matter distribution on Fig.\ref{f0} right consists of two crossed flows, ingoing and outgoing. As a result, the energy flows through $r$-spheres compensate each other and the masses enclosed by $r$-spheres conserve in time. 

The seeming paradox, how black holes can emit matter, is resolved by a detailed calculation in \cite {1701.01569}, which shows that the crossed flows of dark matter erase the event horizon. For the decreasing radius $r$ the enclosed mass $ M (r) $ decreases (imagine how layers of positive mass are successively removed from the star). If the mass decreases sufficiently fast, so that the inequality $ 2GM (r) / c ^ 2 <r $ holds everywhere, then the event horizon is not formed. The calculation shows that this inequality holds for the RDM-star. Thus, RDM-stars are not black holes in the exact sense of the word, they are {\it quasi-black holes}, similar to the ones considered by Visser et al. in \cite{0902.0346}. These are the models where the solution follows the Schwarzschild profile almost to the gravitational radius and then is modified. Because of its T-symmetry, RDM-stars have common features with both black and white holes. As shown in the author's work \cite {1811.03368}, these solutions are directly related to the so-called Eardley's instability of {\it white holes} and are an example of stationary T-symmetric solutions of stable type.

Further, the calculation \cite {1701.01569} shows that crossed flows lead to the phenomenon of mass inflation, considered by Hamilton and Pollack in \cite {0411062}, as a result of which the enclosed mass decreases very fast and becomes negative. In this case, the mass sandwiched between the $r$-spheres grows to huge positive values, and must be compensated by the negative mass located in the center. The author's work \cite {1812.11801} showed that the mass should not be really negative, but can be effectively created by the phenomena of quantum gravity. A similar effect occurs in the {\it Planck stars} model by Rovelli and Vidotto \cite {1401.6562}, where the effective density is written as $ \rho _ {{\rm eff}} = \rho \, (1- \rho / \rho_P) $, whence it is immediately seen that $ \rho> \rho_P $ leads to $ \rho _ {{\rm eff}} <0 $, when the Planck density is exceeded, the gravity turns into its opposite. Thus, a Planck core is formed inside the RDM star, creating the necessary anti-gravity force to balance the enormous mass of the dark matter halo resting on it.

Also, in work \cite {1812.11801}, RDM stars were considered as possible sources of fast radio bursts, arising when the mass of the order of asteroid falls on the Planck core. The resulting ultra-high energy flash is shifted by a superstrong gravitational redshift into the radio band, while calculation \cite {1812.11801} shows that the predicted emission frequency is located within the observed range. The effective negative mass inherent to the Planck core can also lead to the spacetime curvature, characteristic to the {\it wormhole solution}. As a result, a tunnel can be formed leading to another universe or a remote region of ours. Such scenario was calculated in \cite {1707.02764}.

Thus, RDM-stars have a very interesting structure under their gravitational radius, in the region of strong fields. In this paper, however, we will return to the consideration of RDM stars at large distances, in weak fields, and find out whether this model can describe the experimentally observed rotation curves of the galaxies, including their deviations from the simplest flat shape. In Section~2, we compare the predictions of the RDM model with the universal rotation curve describing spiral galaxies of a general form. Section~3 will look at the grand rotation curve, which describes the Milky Way galaxy over a wide range of distances. In Section~4, the modeling of rotation curves at large distances from the center will be considered. In Section~5, we describe the structure of the dark matter density distribution and the corresponding gravitational field outside of the galactic plane.

\section{RDM-stars and Universal Rotation Curve}
Being applied to the galactic disk, the dark-to-luminous matter coupling (\ref{KT1}-\ref{phidm}) produces the following contribution to the orbital acceleration: 
\begin{eqnarray}
&&x=(r_1,0),\ y=(r\cos\alpha,r\sin\alpha),\\
&&a_{r,dm}(x)=G/L_{KT} \int dM_{lm}(y)\, (x-y)_1/(x-y)^2\\
&&=G/L_{KT} \int r dr d\alpha\, \Sigma_{lm}(r)\, (x-y)_1/(x-y)^2\\
&&=G/L_{KT} \int_0^\infty r dr\, \Sigma_{lm}(r)\, I_1(r_1,r),\\
&&I_1(r_1,r)=\int_0^{2\pi}d\alpha\,(r_1-r\cos\alpha)/(r^2+r_1^2-2rr_1\cos\alpha)\\
&&=2\pi/r_1\cdot\theta(r_1-r),\\
&&a_{r,dm}(r_1)=v_{dm}^2(r_1)/r_1= G/(r_1L_{KT}) \int_0^{r_1} 2\pi r dr\, \Sigma_{lm}(r)\\
&&=GM_{lm}(r_1)/(r_1L_{KT}),\ \Sigma_{lm}(r)\sim\exp(-r/R_D),\\ 
&&M_{lm}(r_1)=M_{lm}\cdot(1-\exp(-r_1/R_D) (1 + r_1/R_D)).
\end{eqnarray}
It is necessary to make several comments at this point. It has been noted in \cite{1701.01569}, that the dark matter potential $\varphi_{dm}\sim\log r$ is {\it a fundamental solution} of Poisson equation in 2 dimensions. As a result, the dark matter in-plane gravitational acceleration is described by the enclosed mass function, in the same way as for spherically symmetric mass distribution in 3 dimensions. This property is confirmed by the straightforward computation above, where the internal integral $I_1(r_1,r)$ over a circle is evaluated to the Heaviside $\theta$-function, then the external integral in $a_{r,dm}(r_1)$ is cut at the upper limit $r_1$ and is represented by the luminous matter enclosed mass $M_{lm}(r_1)$. In these formulae $\Sigma_{lm}(r)$ is the luminous mass surface density, for which the exponential Freeman model \cite{freeman} is taken. $M_{lm}(r_1)$ is the corresponding enclosed mass function and the common multiplier $M_{lm}$ without the argument is the total disk mass.

The above described property of the dark matter potential allows to formulate the following set of simple rules for the enclosed mass: 
\begin{itemize}
\item for spherically symmetric distribution of luminous matter, $v_{lm}^2$ contribution is defined in terms of the enclosed luminous mass via Gauss theorem as $v_{lm}^2=GM_{lm}(r)/r$;
\item for dark matter coupled to luminous matter via KT-integral and planar axially symmetric distribution of the luminous matter, in-plane $v_{dm}^2$ contribution is defined in terms of the enclosed luminous mass as $v_{dm}^2=GM_{lm}(r)/L_{KT}$;
\item asymptotically, for compact distribution of luminous matter at large distances, $v_{lm}^2$ and $v_{dm}^2$ contributions are defined in terms of the total enclosed luminous mass as $v_{lm}^2=GM_{lm}/r$ and $v_{dm}^2=GM_{lm}/L_{KT}$. 
\end{itemize}
In the other cases, e.g., for exponential disk at moderate distances, the mass distribution is not spherically symmetric and $v_{lm}^2$ contribution is not defined in terms of the enclosed mass by a local formula.

Note also that in paper \cite{0604496} the dark-to-luminous matter coupling has been introduced in the form (\ref{KT1}-\ref{KT2}), then it has been modified by an oscillatory term. The main conclusions of \cite{0604496} are preserved without this oscillatory term and we will also consider the version of KT-integral without the oscillatory term in this paper. 

The above computed dark matter profile together with the formulae \cite{freeman,9506004} for luminous matter allow to formulate the following model:
\begin{eqnarray}
&&v^2_{center,lm}=\alpha_0 v_{opt}^2\, R_{opt}/r,\ v^2_{center,dm}=\alpha v_{opt}^2,\\
&&v^2_{disk,lm}=\beta v_{opt}^2\, F_{disk}(r/R_D)/F_{disk}(3.2),\\
&&F_{disk}(x)=x^2(I_0(x/2) K_0(x/2)-I_1(x/2) K_1(x/2)),\\ 
&&v^2_{disk,dm}=\gamma v_{opt}^2\, F_{disk,dm}(r/R_D)/F_{disk,dm}(3.2),\\
&&F_{disk,dm}(x)=1-e^{-x}(1+x),\ R_{opt}=3.2R_D,\\
&&v^2=v^2_{center,lm}+v^2_{center,dm}+v^2_{disk,lm}+v^2_{disk,dm},\\
&&v^2_{opt}=v^2(r\to R_{opt}),\ \alpha_0+\alpha+\beta+\gamma=1,
\end{eqnarray}
where $I_n,K_n$ are the modified Bessel functions. We took into account the contribution of the galactic center, including the central supermassive black hole and the bulge. At the considered range of distances it is represented by an unresolved point-like contribution and is described by the asymptotic compact mass formulae above. The velocities are normalized to $R_{opt}$ distance, under which 83\% of the luminous mass is located. The coefficients in the additive formulae for squared velocities become the parameters for fitting the model to the experimental data, which we will now perform.

\begin{figure}
\begin{center}
\includegraphics[width=0.7\textwidth]{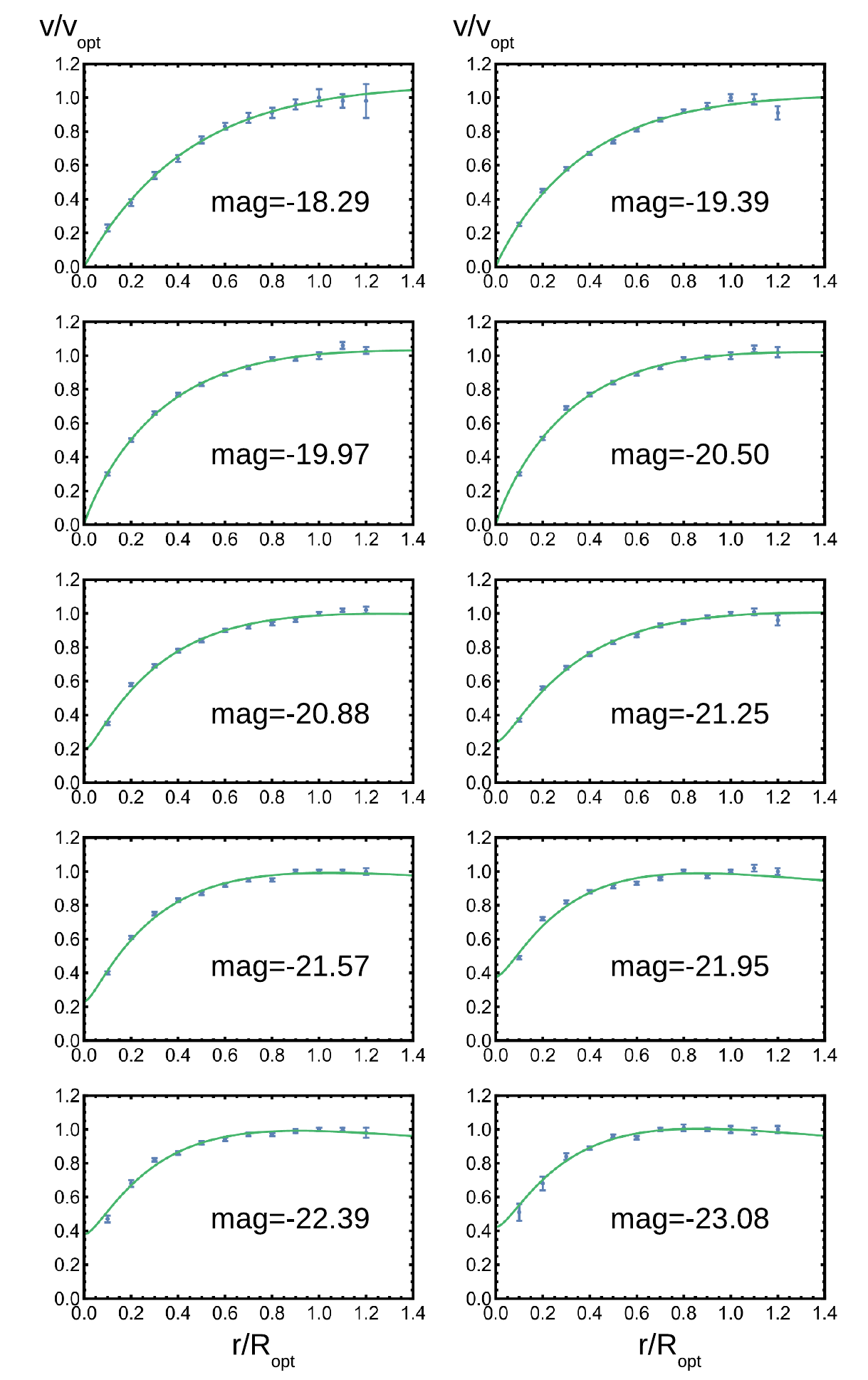}
\end{center}
\caption{RDM model fits the experimental rotation curves. Data from \cite{9502091}.}\label{f3}
\end{figure}

\begin{figure}
\begin{center}
\includegraphics[width=0.7\textwidth]{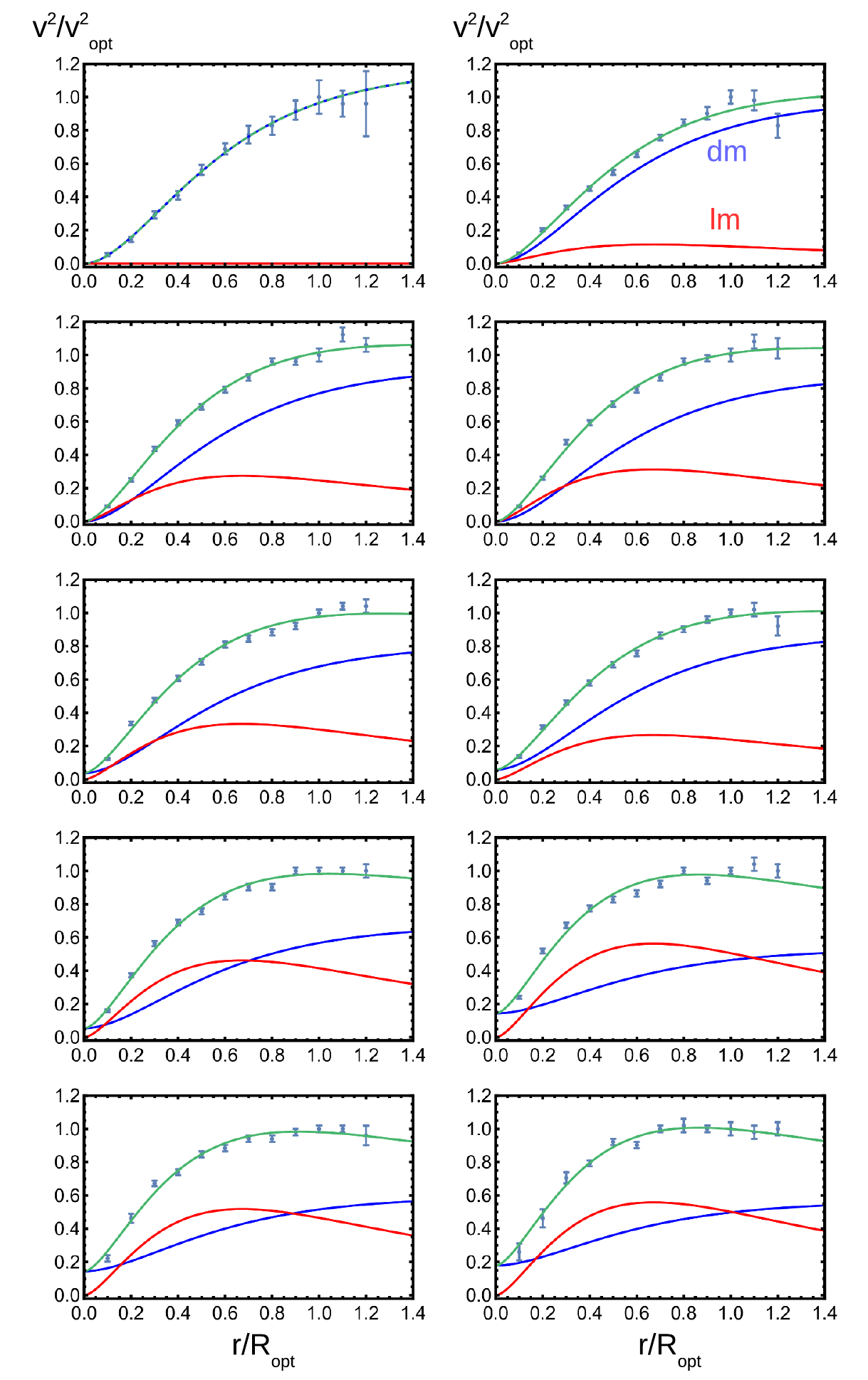}
\end{center}
\caption{The same plots in velocity squared coordinate. The contributions of luminous matter (red) and dark matter (blue) are separated. The green line is the sum of these contributions.}\label{f3b}
\end{figure}

\begin{figure}
\begin{center}
\includegraphics[width=\textwidth]{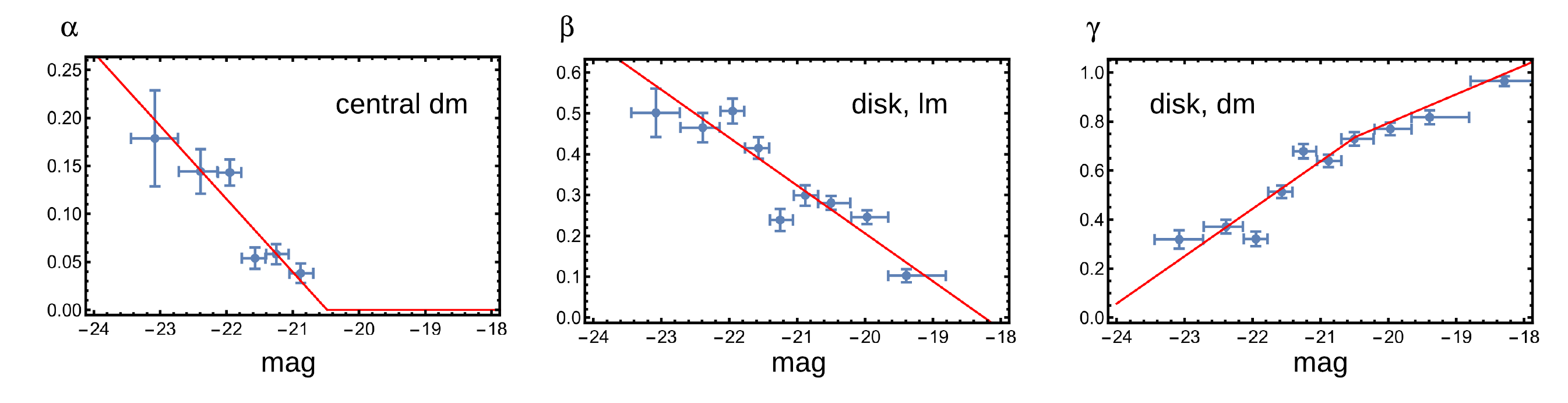}
\end{center}
\caption{The fit of $\alpha$, $\beta$, $\gamma$ coefficients as functions of the magnitude. The contributions of central dark matter, disk luminous matter and disk dark matter are separated.}\label{f4}
\end{figure}

The data \cite{9502091} represent the rotation curves of 967 spiral galaxies, normalized to $r/R_{opt}$ and $v/v_{opt}$, averaged in 10 luminosity bins. The same data have been used later in \cite{9503051} to evaluate the dark matter contribution. Then in \cite{9506004} they have been extended by an additional dataset and the behavior of the rotation curve in the range $1\leq r/R_{opt}\leq 2$ has been estimated. Further, in \cite{0703115} even larger distances have been considered. In this section we restrict our analysis to the original \cite{9502091} dataset with the range $0.1\leq r/R_{opt}\leq 1.2$ and discuss the modeling of larger distances in further sections. 

The experimental data are shown in Fig.\ref{f3} by blue points with error bars. We see that the model, represented by green lines, fits the data perfectly. 
On Fig.\ref{f3b} we plot the same data in velocity squared coordinate and show the separated contributions of luminous matter (red) and dark matter (blue). The dark matter central and disk contributions on this plot are added together: $v^2_{dm}(r)=v^2_{center,dm}+v^2_{disk,dm}(r)$. To separate them back, one can define a constant $v^2_{center,dm}=v^2_{dm}(0)$ and subtract it to obtain the pure disk contribution: $v^2_{disk,dm}(r)=v^2_{dm}(r)-v^2_{dm}(0)$. We remind that all contributions are additive on the velocity squared plot.

Considering this plot further, for small luminosity ($mag=-18.29$) dark matter describes the experimental rotation curves dominantly, with a negligible luminous matter contribution. Then, with increasing luminosity, the contribution of dark matter diminishes, while the contribution of luminous matter increases. For higher luminosity ($mag=-23.08$) the contribution of luminous matter starts to prevail at intermediate distances, while at large distances the dark matter dominates always. 

This behavior qualitatively coincides with \cite{9506004}. The difference is the other form of dark matter term:
\begin{eqnarray}
&&v^2_{disk,dm}\sim(r/R_{opt})^2/((r/R_{opt})^2+a^2),
\end{eqnarray}
which represents {\it the cored} dark matter distribution with the core radius $a$ becoming in \cite{9506004} a function of luminosity. Such velocity distribution is produced by a spherically symmetric halo with a bounded density in the center $\rho_{sph}(0)<C$ and enclosed mass function $M_{sph}(r)\sim r^3$, so that the squared velocity becomes $v_{sph}^2=GM_{lm}(r)/r\sim r^2$. Note that in our model the dark matter disk contribution has the same behavior. Specifically, the dark matter contribution $v^2_{disk,dm}=GM_{lm}(r)/L_{KT}$ is proportional to the planar enclosed mass function $M_{lm}(r)\sim r^2$ and possesses the same $r\to0$ asymptotics as $v_{sph}^2$. 

On the other hand, our dark mass density distribution is different from the spherically symmetric cored one. The luminous matter is located in a plane $\rho_{lm}\sim\delta(z)$, while the dark matter is coupled to it by KT-integral. This integral is not spherically symmetric and, as we see in the next sections, contains a similar $z$-singularity. Nevertheless, the corresponding dark mass contribution to the in-plane squared velocity near the center is {\it the same} as for the cored distribution.

The other difference from \cite{9506004} is the central term $v^2_{center,dm}=Const$ activated at higher luminosity bins. It corresponds to the non-cored distribution with the original unbounded $\rho_{center,dm}\sim r^{-2}$ dependence. This dependence is valid as long as the center remains unresolved, while for the distances in the bulge region, it starts to show the cored behavior. Only the dark matter contribution of supermassive black hole in the center of the galaxy remains unresolved till its gravitational radius, however, it does not prevail over the other central structures in the model. In the next section we will follow the distance scale into the bulge and consider this question in more detail.

\begin{table}
\begin{center}
\caption{URC fitting results}\label{tab4}

~

\def\arraystretch{1.1}
\begin{tabular}{|c|c|c|c|}
\hline
mag&$\alpha$, central dm &$\beta$, disk lm &$\gamma$, disk dm \\
\hline
-18.29  & 0 & 0 &  0.965 $\pm$ 0.020\\
-19.39  & 0  &  0.103 $\pm$ 0.016 &  0.817 $\pm$ 0.028\\
-19.97  & 0   &  0.246 $\pm$ 0.017 &  0.770 $\pm$ 0.026\\
-20.5   & 0  &  0.281 $\pm$ 0.017 &  0.729 $\pm$ 0.027\\
-20.88  &  0.038 $\pm$ 0.010  &  0.299 $\pm$ 0.025 &  0.640 $\pm$ 0.025\\
-21.25  &  0.058 $\pm$ 0.011  &  0.239 $\pm$ 0.027  &  0.679 $\pm$ 0.029\\
-21.57  &  0.054 $\pm$ 0.011  &  0.415 $\pm$ 0.026  &  0.513 $\pm$ 0.026\\
-21.95  &  0.143 $\pm$ 0.013  &  0.505 $\pm$ 0.030  &  0.321 $\pm$ 0.030\\
-22.39  &  0.144 $\pm$ 0.023  &  0.465 $\pm$ 0.036  &  0.371 $\pm$ 0.028\\
-23.08  &  0.179 $\pm$ 0.050  &  0.501 $\pm$ 0.059  &  0.319 $\pm$ 0.038\\
\hline
\end{tabular}

\end{center}
\end{table}

\begin{table}
\begin{center}
\caption{Linear fit for URC shape coefficients*}\label{tab5}

~

\def\arraystretch{1.1}
\begin{tabular}{|c|c|c|}
\hline
&$\alpha$&$\beta$\\
\hline
$c_0$ & -1.56 $\pm$ 0.24 & -2.14 $\pm$ 0.17 \\
$c_1$ & -0.076 $\pm$ 0.011 & -0.117 $\pm$ 0.008 \\
\hline
\end{tabular}

\vspace{2mm}
{\footnotesize * $\alpha=\max(c_{\alpha0}+c_{\alpha1}\,mag,0)$, $\beta=c_{\beta0}+c_{\beta1}\,mag$, $\gamma=1-\alpha-\beta$}

\end{center}
\end{table}

In the rest of this section we will provide technical details about the used fitting procedure. At first, we fitted the experimental curves by all four basis forms including the contribution of central luminous matter with the coefficient $\alpha_0$. As a result, we obtained the values $\alpha_0$ comparable with zero. Further, we set $\alpha_0=0$ and repeated the fit with three basis forms. At the first four luminosity bins the obtained values for the coefficient $\alpha$, regulating the contribution of central dark matter, was negative or comparable with zero up to an error. Also in the first luminosity bin the value $\beta$ for the contribution of the luminous disk was comparable with zero. In those cases the values of the corresponding coefficients have been set to zero, and the fit was repeated with a reduced number of basis shapes. The obtained results are placed in Table~\ref{tab4}. The comparable results are also obtained in the following versions of the fitting procedure:
\begin{itemize}
\item the non-negativity of shape coefficients explicitly written as inequality constraints;
\item the exponential substitutions done for the coefficients, $par=\exp(lpar)$, in this case the coefficient becoming fixed to zero in the first approach, results in large negative $lpar$, so that the coefficient is effectively nullified.
\end{itemize}

The constraint $\alpha+\beta+\gamma=1$ has not been imposed during the fit. However, since the experimental data have been normalized in the point $r/R_{opt}=1$ to $v/v_{opt}=1$ and the fitted curves follow this normalization closely, this constraint has been fulfilled up to an error on the fitting results. Then, $\alpha$ and $\beta$ coefficients have been linearly fitted with respect to $mag$, with the necessary clamp to avoid the negativity, see Fig.\ref{f4} and Table~\ref{tab5}, and the remaining coefficient $\gamma$ has been found from normalization:
\begin{eqnarray}
&&\alpha=\max(-1.56-0.076\,mag,0),\\ 
&&\beta=-2.14-0.117\,mag,\ \gamma=1-\alpha-\beta,\\ 
&&-23.08\leq mag\leq-18.29.
\end{eqnarray}

\section{RDM-stars and Rotation Curve for Milky Way}
\begin{figure}
\begin{center}
\includegraphics[width=0.45\textwidth]{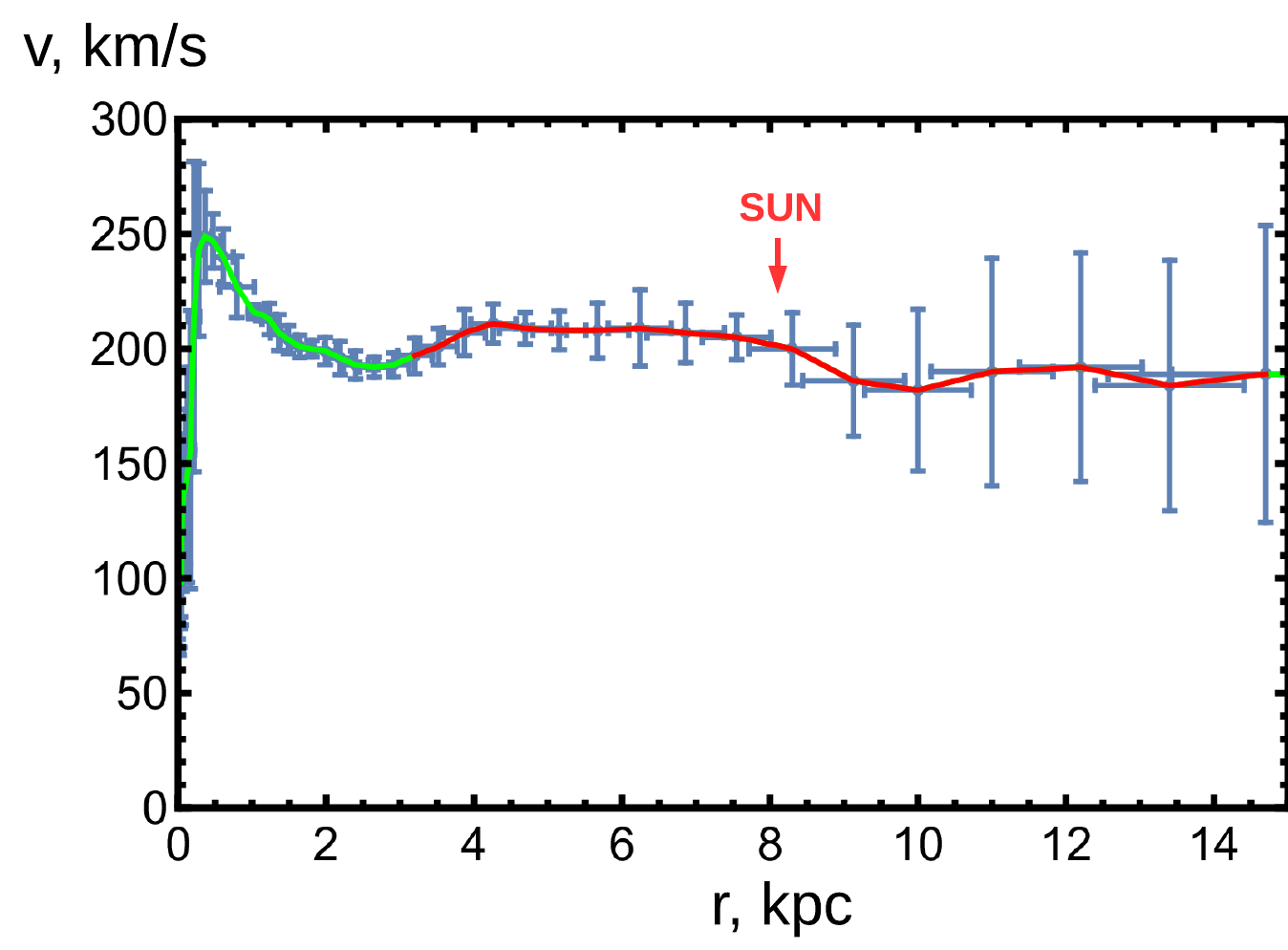}
\includegraphics[width=0.45\textwidth]{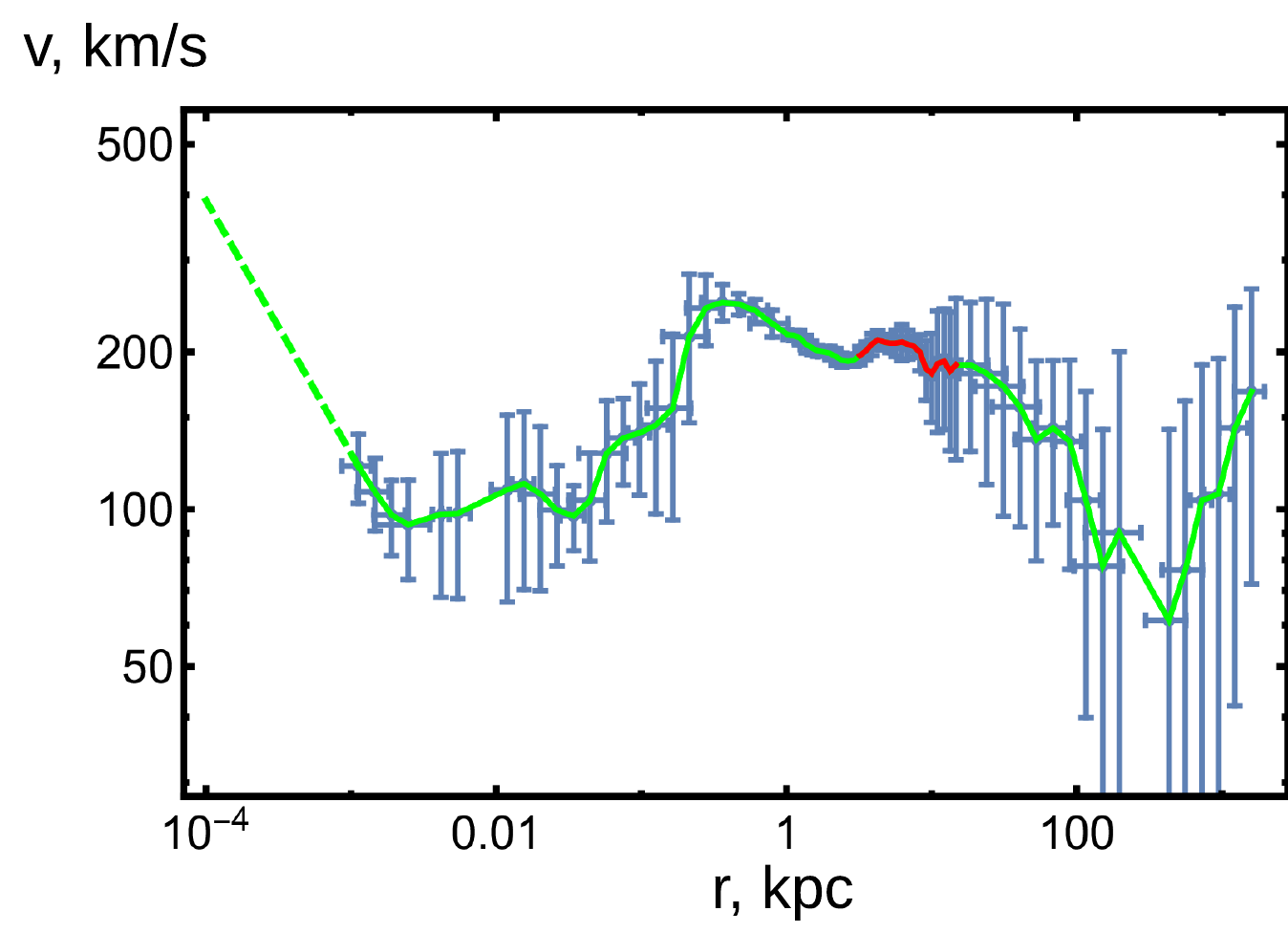}
\end{center}
\caption{Grand rotation curve in linear and logarithmic scale. Data from~\cite{1307.8241}.}\label{f1}
\end{figure}

In the paper \cite{1307.8241} the experimental rotation curve for the entire Milky Way galaxy has been presented. This curve, continuously covering the range of distances from the Galactic Center to the size of the Local Group, is shown on Fig.\ref{f1}. In linear scale on the left there is a large segment of the curve in the range 3-15kpc that looks approximately flat. In logarithmic scale on the right one can see more complicated structures: the minimum of the curve beyond 100kpc, where the dark halo ends, two maxima near 0.5kpc and 0.01kpc, corresponding to the main bulge and the inner bulge, in terminology of \cite{1307.8241}, and the central 1pc zone of Keplerian domination of the supermassive black hole, shown by the dashed line.

\begin{figure}
\begin{center}
\includegraphics[width=0.45\textwidth]{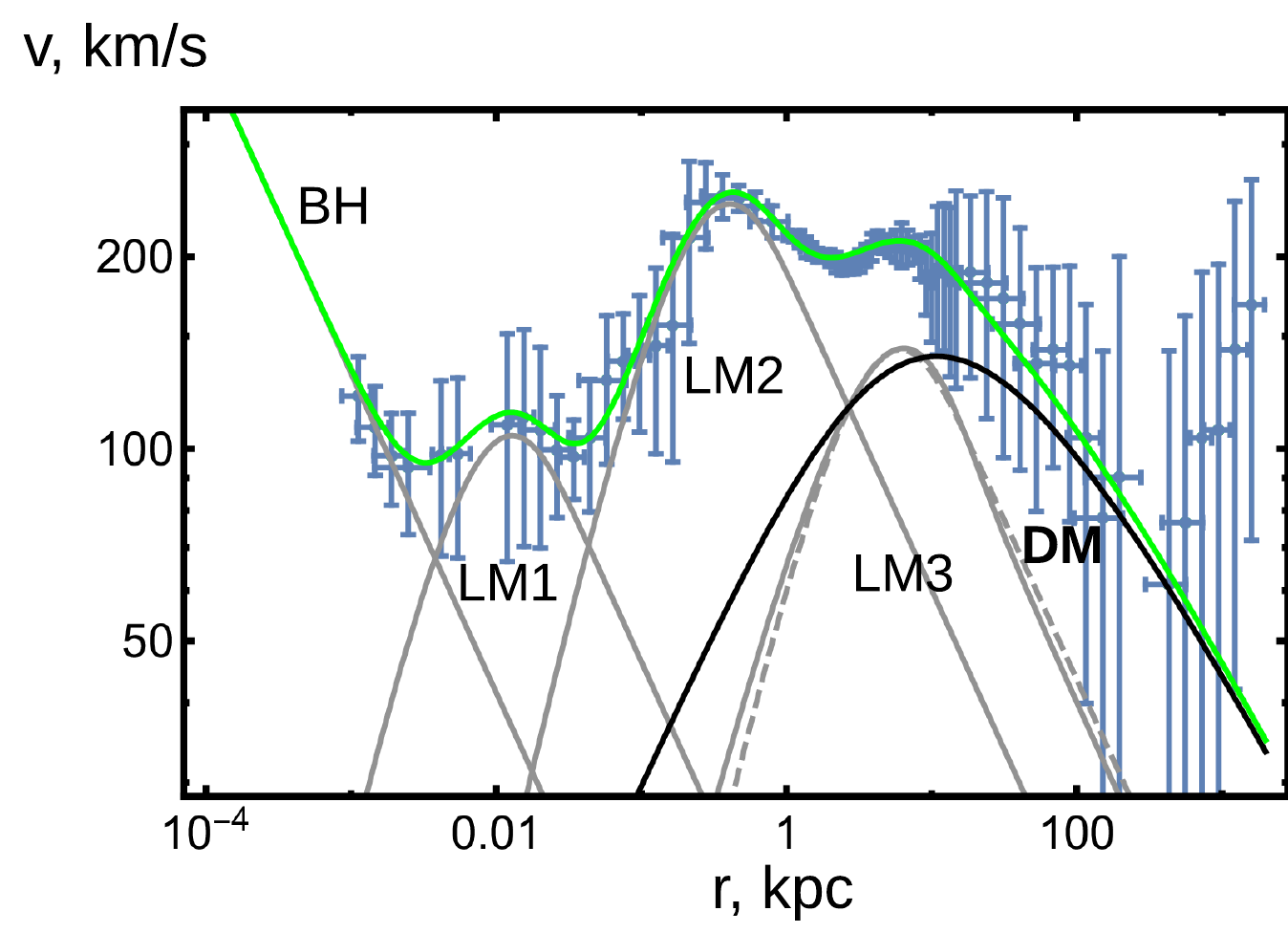}
\includegraphics[width=0.45\textwidth]{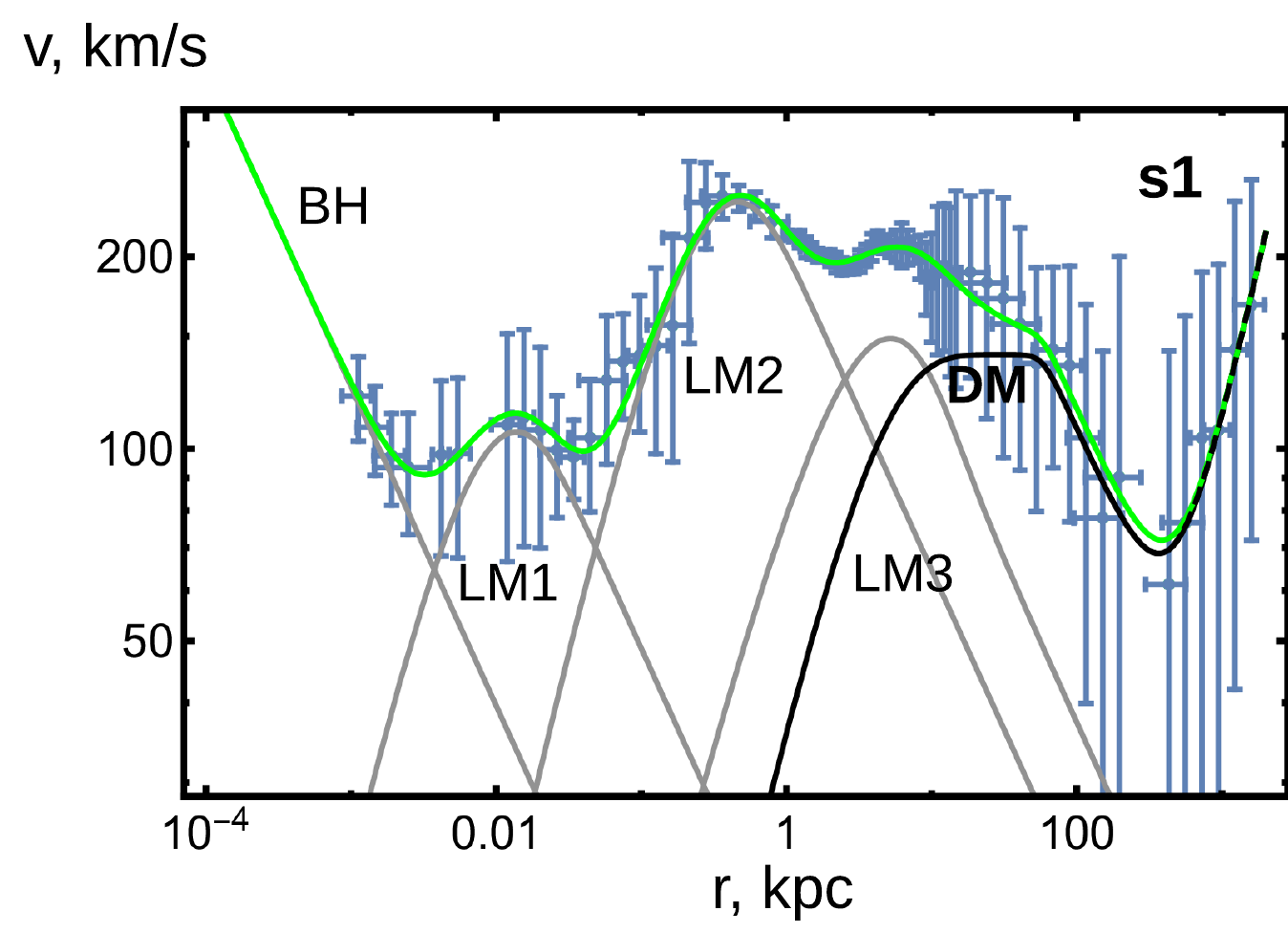}

\includegraphics[width=0.45\textwidth]{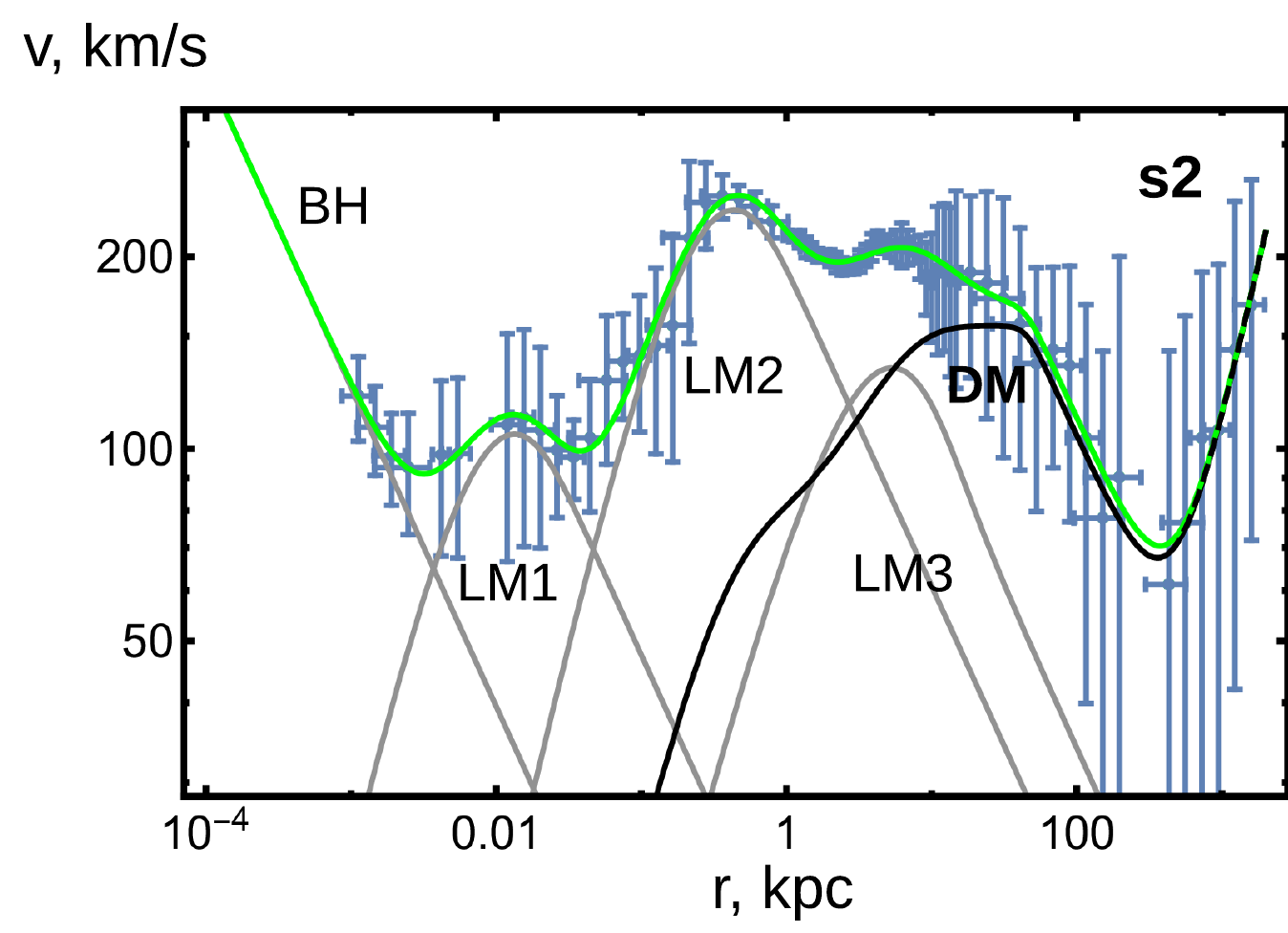}
\includegraphics[width=0.45\textwidth]{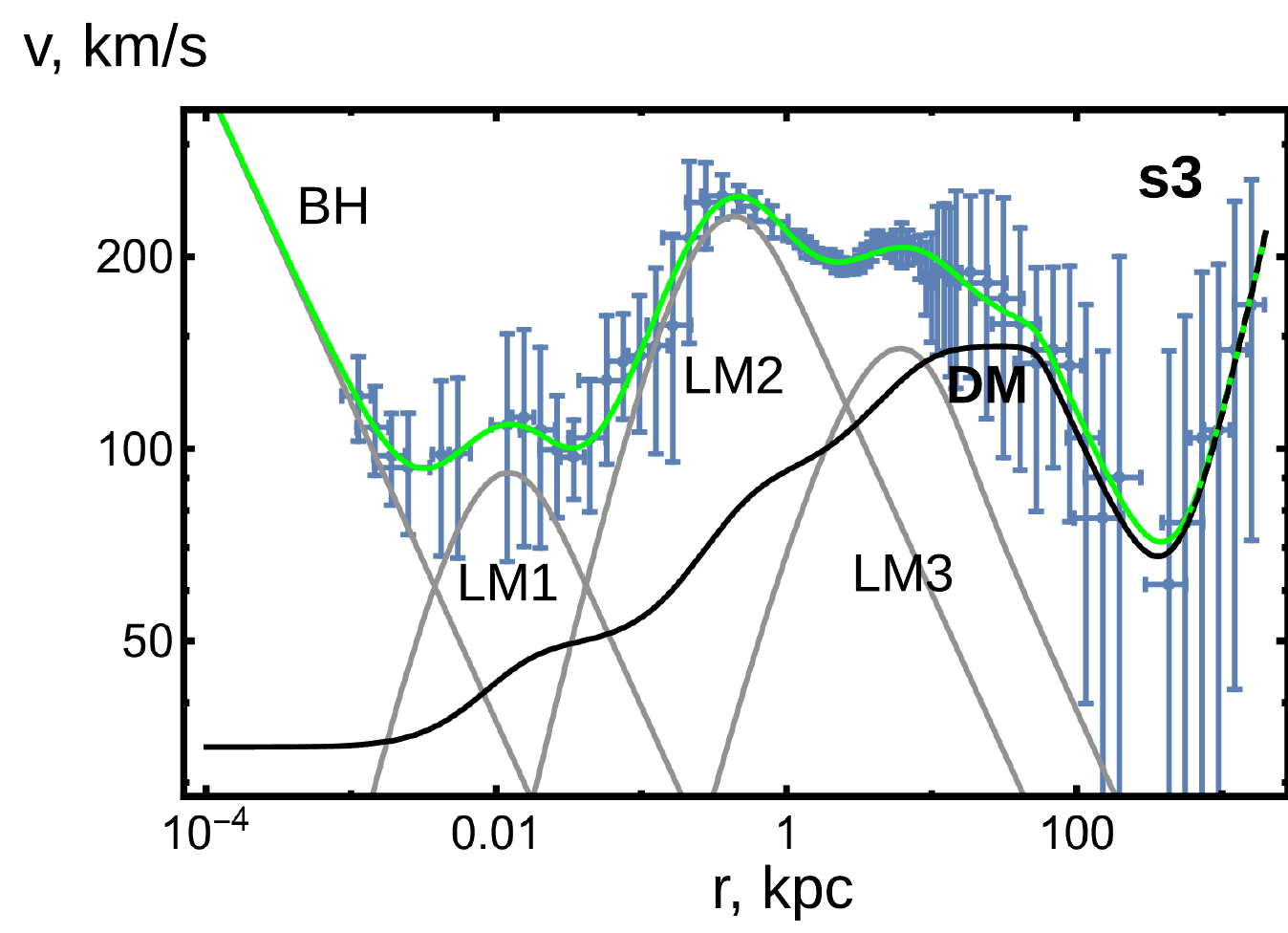}
\end{center}
\caption{Model fits of the grand rotation curve. Top left -- the fit \cite{1307.8241} with NFW DM profile; other plots -- RDM model fits for three coupling scenarios. BH is the Keplerian contribution of the central black hole; LM1-3 are contributions of luminous matter from the inner bulge, the main bulge and the disk; DM is the dark matter contribution. The green line is the quadratic sum of these contributions.}\label{f2}
\end{figure}

For the fit we need to evaluate KT-integral for the distribution of the luminous matter in the bulge components given by the {\it exponential spheroid model} \cite{1307.8241}. This is done by a straightforward computation:
\begin{eqnarray}
&&x=(r_1,0,0),\ y=(r\cos\alpha,r\sin\alpha,0),\\
&&a_r(x)=G/L_{KT} \int dM_{lm}(y)\, (x-y)_1/(x-y)^2\\
&&=G/L_{KT} \int r^2 dr\, 2\pi\sin\alpha\, d\alpha\, \rho_{lm}(r)\, (x-y)_1/(x-y)^2\\
&&=G/L_{KT} \int_0^\infty 2\pi r^2 dr \rho_{lm}(r)\, I_2(r_1,r),\\
&&I_2(r_1,r)=\int_0^{\pi} d\alpha\,\sin\alpha\,(r_1-r\cos\alpha)/(r^2+r_1^2-2rr_1\cos\alpha)\\
&&=\int_{-1}^1 du\,(r_1-ru)/(r^2+r_1^2-2rr_1u)\\
&&=(2 r r_1 + (r^2 - r^2_1)\log(|r - r_1|/(r + r_1)))/( 2 r r_1^2),\\
&&\rho_{lm}(r)= M_{lm}/(8\pi a^3)\cdot\exp(-r/a),\\ 
&&a_r(r_1)=GM_{lm}/(4 a^3 L_{KT})\int_0^\infty r^2 dr \exp(-r/a)\, I_2(r_1,r)\\
&&=GM_{lm}/(r_1 L_{KT})\cdot I_3(r_1/a),\\
&&I_3(x)=((3-3x+x^2)e^xEi(-x)-(3+3x+x^2)e^{-x}Ei(x)\\
&&+6x)/(4x).
\end{eqnarray}
Unifying it with the luminous matter contributions from \cite{1307.8241}, we obtain the following model:
\begin{eqnarray}
&&v^2_{bh}=G_m M_{bh}/r,\ G_m=4.3016\cdot10^{-6}~(km/s)^2(kpc/M_\odot),\\
&&v^2_{sph,i}=G_m M_i/r \cdot F_{sph}(r/a_i),\ i=1,2,\\
&&F_{sph}(x)=1-e^{-x}(1+x+x^2/2),\\
&&v^2_{disk}=G_m M_{disk}/(2 R_D)\cdot F_{disk}(r/R_D),\\
&&F_{disk}(x)=x^2(I_0(x/2) K_0(x/2)-I_1(x/2) K_1(x/2)),\label{Fdisk}\\ 
&&v^2_{lm}=v^2_{bh}+v^2_{sph1}+v^2_{sph2}+v^2_{disk},\\
&&v^2_{dm,bh}=G_m M_{bh}\lambda_{bh}/L_{KT},\\
&&v^2_{dm,sph,i}=G_m M_i\lambda_{sph,i}/L_{KT}\cdot F_{dm,sph}(r/a_i),\ i=1,2,\\
&&F_{dm,sph}(x)=(6x+(3 - 3x + x^2) e^{x} Ei(-x) \\
&&\quad - (3 + 3x + x^2) e^{-x} Ei(x))/(4x),\\
&&v^2_{dm,disk}=G_m M_{disk}\lambda_{disk}/L_{KT}\cdot F_{dm,disk}(r/R_D),\\
&&F_{dm,disk}(x)=1-e^{-x}(1+x),\\
&&v^2_{dm,sum}=v^2_{dm,bh}+v^2_{dm,sph1}+v^2_{dm,sph2}+v^2_{dm,disk},\\
&&v^2_{dm,cut}=v^2_{dm,sum}(r\to r_{cut})\cdot r_{cut}/r,\ 
v^2_{bgr}=4\pi G_m \rho_0 r^2/3,\\
&&v^2_{dm}=((v^2_{dm,sum})^p+(v^2_{dm,cut})^p)^{1/p},\\
&&v^2=v^2_{lm}+v^2_{dm}+v^2_{bgr}.
\end{eqnarray}
Here $I_n,K_n$ are the modified Bessel functions, $Ei$ is the exponential integral function. $G_m$ represents the gravitational constant $G$ in the model system of units, measuring masses in $M_\odot$, lengths in kpc, velocities in km/s. 

The features of the rotation curve at large distances are simulated as follows. We perform a cut of dark matter density on the outer distance $r_{cut}$, to obtain a halo of a finite radius. At larger distances the rotation curve switches to the Keplerian regime $v^2_{dm,cut}(r)$. A similar cut has been also performed in \cite{0505131}. Then, to join the increasing function $v^2_{dm,sum}(r)$ and the decreasing function $v^2_{dm,cut}(r)$ we take $v^2_{dm}=\min(v^2_{dm,sum},v^2_{dm,cut})$, define it as $L_{-\infty}$-norm, then replace it by a smooth $L_p$-norm with a numerical value $p=-10$. These manipulations on the rotation curve have negligible influence to the fitting procedure, it is only important that a flat or slightly increasing dark matter profile is changed to the falling profile at large distances. We also add a uniform background mass density contribution $v^2_{bgr}$ to simulate the increase of the rotation curve after the minimum. The other possibilities for modeling of the outer part of the rotation curve will be discussed in the next section. 

We introduce the dimensionless coefficients $\lambda_i$, defining the coupling of dark matter to different components of the luminous matter. This allows to take into account a different concentration or a different population of the black holes in different components. The resulting dark matter contributions (bh,sph,disk) have a common form, $G_m M_i\lambda_i/L_{KT}$ multiplied to a function of $r$, tending to unity when $r\to\infty$. If the cutting is performed at sufficiently large $r_{cut}$, the total enclosed mass of dark matter is
\begin{eqnarray}
&&M_{dm}(r_{cut})=v^2_{dm,sum}(r_{cut})r_{cut}/G_m=r_{cut}/L_{KT}\cdot \\
&&\quad (\lambda_{bh} M_{bh}+\lambda_{sph1} M_1+\lambda_{sph2} M_2+\lambda_{disk} M_{disk}).
\end{eqnarray}

The results of the fit are presented in Fig.\ref{f2}. The top left image shows the fit of \cite{1307.8241}, where the NFW profile has been taken to represent the dark matter halo. The BH part corresponds to the Keplerian contribution of the central black hole; LM1 and LM2 are the luminous mass spheroidal contributions of the inner and the main bulge. The LM3 solid line corresponds to the luminous mass disk contribution, while the dashed line shows a spheroidal contribution close to it. These two contributions look similar, but differ in details. In particular, they possess slightly different asymptotics at $r\to0$, $v^2_{sph}\sim r^2$ vs $v^2_{disk}\sim r^2(-\log r)$. The last formula is obtained by decomposing the Bessel functions in (\ref{Fdisk}). The velocity squared sum shown by the green line fits the experimental data perfectly.

The next three images in Fig.\ref{f2} show the fit of the same data by the model constructed in this paper, for three different selections of $\lambda_i$ coefficients shown in Table~\ref{tab1}. As we see, they fit the experimental data perfectly as well. This means that the fit of the data can neither favor NFW/RDM profile nor decide between RDM scenarios with different couplings. A possible reason is that the rotation curve of the Milky Way seems to be dominated by the luminous matter, while the dark matter becomes important only at the large distances, where the data scatter is so large. As a result, quite different dark matter profiles can describe the same data equally good.

\begin{table}
\begin{center}
\caption{GRC: fixed DM coupling coefficients for 3 scenarios}\label{tab1}

~

\def\arraystretch{1.1}
\begin{tabular}{|c|c|c|c|}
\hline
$\lambda_{KT}$&s1&s2&s3\\
\hline
 $\lambda_{bh}$& 0& 1& $10^3$ \\
 $\lambda_1$& 0& 1& $10^2$ \\
 $\lambda_2$& 0& 1& 2 \\
 $\lambda_{disk}$& 1& 1& 1 \\
\hline

\end{tabular}

\end{center}
\end{table}

\begin{table}
\begin{center}
\caption{GRC: fitting results, log of parameters and their errors*}\label{tab2}

~

\def\arraystretch{1.1}
\begin{tabular}{|c|c|c|c|}
\hline
$\log(par)$&s1&s2&s3\\
\hline
 $M_{bh}$& 15.1 $\pm$ 0.2& 15.1 $\pm$ 0.2& 15.0 $\pm$ 0.3 \\
 $M_1$& 17.8 $\pm$ 0.3& 17.8 $\pm$ 0.3& 17.4 $\pm$ 0.5 \\
 $a_1$& -5.5 $\pm$ 0.4& -5.5 $\pm$ 0.4& -5.6 $\pm$ 0.4 \\
 $M_2$& 22.99 $\pm$ 0.04& 22.88 $\pm$ 0.13& 22.83 $\pm$ 0.13 \\
 $a_2$& -1.99 $\pm$ 0.07& -2.05 $\pm$ 0.10& -2.05 $\pm$ 0.09 \\
 $M_{disk}$& 24.2 $\pm$ 0.8& 24.0 $\pm$ 0.8& 24.3 $\pm$ 0.5 \\
 $R_D$& 0.9 $\pm$ 0.3& 0.91 $\pm$ 0.17& 1.04 $\pm$ 0.11 \\
 $L_{KT}$& 1.9 $\pm$ 1.4& 1.8 $\pm$ 1.1& 2.5 $\pm$ 0.8 \\
 $r_{cut}$& 4.1 $\pm$ 0.9& 3.8 $\pm$ 0.7& 4.0 $\pm$ 0.7  \\
 $\rho_0$& 6.5 $\pm$ 0.8& 6.5 $\pm$ 0.7& 6.5 $\pm$ 0.8 \\
\hline

\end{tabular}

{\footnotesize * masses in $M_\odot$, lengths in $kpc$, density in $M_\odot/kpc^3$}

\end{center}
\end{table}

\begin{table}
\begin{center}
\caption{GRC: fitting results, central values of parameters*}\label{tab3}

~

\def\arraystretch{1.1}
\begin{tabular}{|c|c|c|c|}
\hline
$par$&s1&s2&s3\\
\hline
 $M_{bh}$& $3.6\times 10^6$ & $3.6\times 10^6$ & $3.2\times 10^6$ \\
 $M_1$& $5.5\times 10^7$ & $5.2\times 10^7$ & $3.6\times 10^7$ \\
 $a_1$& $0.0041$ & $0.0039$ & $0.0036$ \\
 $M_2$& $9.7\times 10^9$ & $8.6\times 10^9$ & $8.2\times 10^9$ \\
 $a_2$& $0.13$ & $0.13$ & $0.13$ \\
 $M_{disk}$& $3.2\times 10^{10}$ & $2.7\times 10^{10}$ & $3.5\times 10^{10}$ \\
 $R_D$& $2.4$ & $2.5$ & $2.8$ \\
 $L_{KT}$& $7.0$ & $6.3$ & $12.0$ \\
 $r_{cut}$& $58$ & $45$ & $53$ \\
 $M_{dm}(r_{cut})$& $2.7\times 10^{11}$ & $2.5\times 10^{11}$ & $2.6\times 10^{11}$ \\
 $\rho_0$& $646$ & $653$ & $649$ \\
\hline

\end{tabular}

{\footnotesize * masses in $M_\odot$, lengths in $kpc$, density in $M_\odot/kpc^3$}

\end{center}
\end{table}

Here we provide the technical details on the fitting procedure. We perform the exponential substitution $par=\exp(lpar)$ and select the logarithms of all free variables as fitting parameters. Table~\ref{tab2} presents the resulting values of these logarithms together with their standard errors. In Table~\ref{tab3} we present the central values of parameters in the physical units. 

The mass of the central black hole could be constrained to a more precise value, e.g., $M_{bh}=(4.1\pm0.6)\cdot10^6 M_\odot$ by Ghez et al. \cite{0808.2870}, however its influence on the fit is negligible, and its logarithm $\log(M_{bh}/M_\odot)\sim15.2$ comparing with our fit results $15.1\pm0.2$, $15.0\pm0.3$ is in the range of one standard deviation. 

The resulting composition of the galaxy shows the strong hierarchy of masses and distances: 
\begin{eqnarray}
&&M_{bh}\ll M_1\ll M_2\ll M_{disk}\ll M_{dm},\\
&&a_1\ll a_2\ll R_D\ll r_{cut}.
\end{eqnarray}
The coupling length $L_{KT}$ varies between the scenarios and is of the order of $R_{opt}=3.2R_D$. 

The baryon fraction $M_{lm}/(M_{lm}+M_{dm}(r_{cut}))$ for the data in Table~\ref{tab3} varies in the range 0.12-0.14. It is located between the estimation 0.12 for the groups of galaxies \cite{1110.4431,1004.2785} and the cosmological value 0.17 from WMAP \cite{0803.0586}, corrected later to 0.16 by Planck observations \cite{1807.06209}. 

The background density $\rho_0\sim 650 M_\odot/kpc^3$ is approximately 5 times greater than the critical density $\rho_{c0}=3H_0^2/(8\pi G)\sim 126 M_\odot/kpc^3$, evaluated for $H_0=67.4~km/s/Mpc$ \cite{1807.06209}. This result indicates the background mass overdensity in the Local Group in comparison with the cosmological value. 

In more detail, assuming that the background density is composed of the uniform baryonic, dark matter and dark energy contributions in the Local Group, we have: $\rho_0=\rho_{bLG}+\rho_{dmLG}-2\rho_\Lambda$. Here we took the effective negative contribution for dark energy term (see, e.g., Blau \cite{Blau}). Thus, we have $\rho_0/\rho_{c0}=\Omega_{bLG}+\Omega_{dmLG}-2\Omega_{\Lambda0}$. Using the known $\Omega_{\Lambda0}=0.68$ \cite{1807.06209} and the observed $\rho_0/\rho_{c0}\sim5$, for the local overdensity of the baryonic and dark matter contribution relative to the cosmological value we obtain: 
\begin{eqnarray}
&(\Omega_{bLG}+\Omega_{dmLG})/(\Omega_{b0}+\Omega_{dm0})=(\rho_0/\rho_{c0}+2\Omega_{\Lambda0})/(1-\Omega_{\Lambda0})\sim20.
\end{eqnarray}

\section{The outer part of the rotation curve}

The modeling of the outer rotation curve in the dark matter halo uses Navarro-Frenk-White (NFW) profile \cite{1307.8241} or Burkert profile \cite{0703115}, which has the same outer asymptotics as NFW one. The NFW profile originates from {\it numerical N-body simulations} of galaxies. Note that there are many other profiles used to describe the simulations. The paper by Merritt et al. \cite{0509417} argues that Einasto profile describes the simulations better than NFW one.

\begin{figure}
\begin{center}
\includegraphics[width=0.45\textwidth]{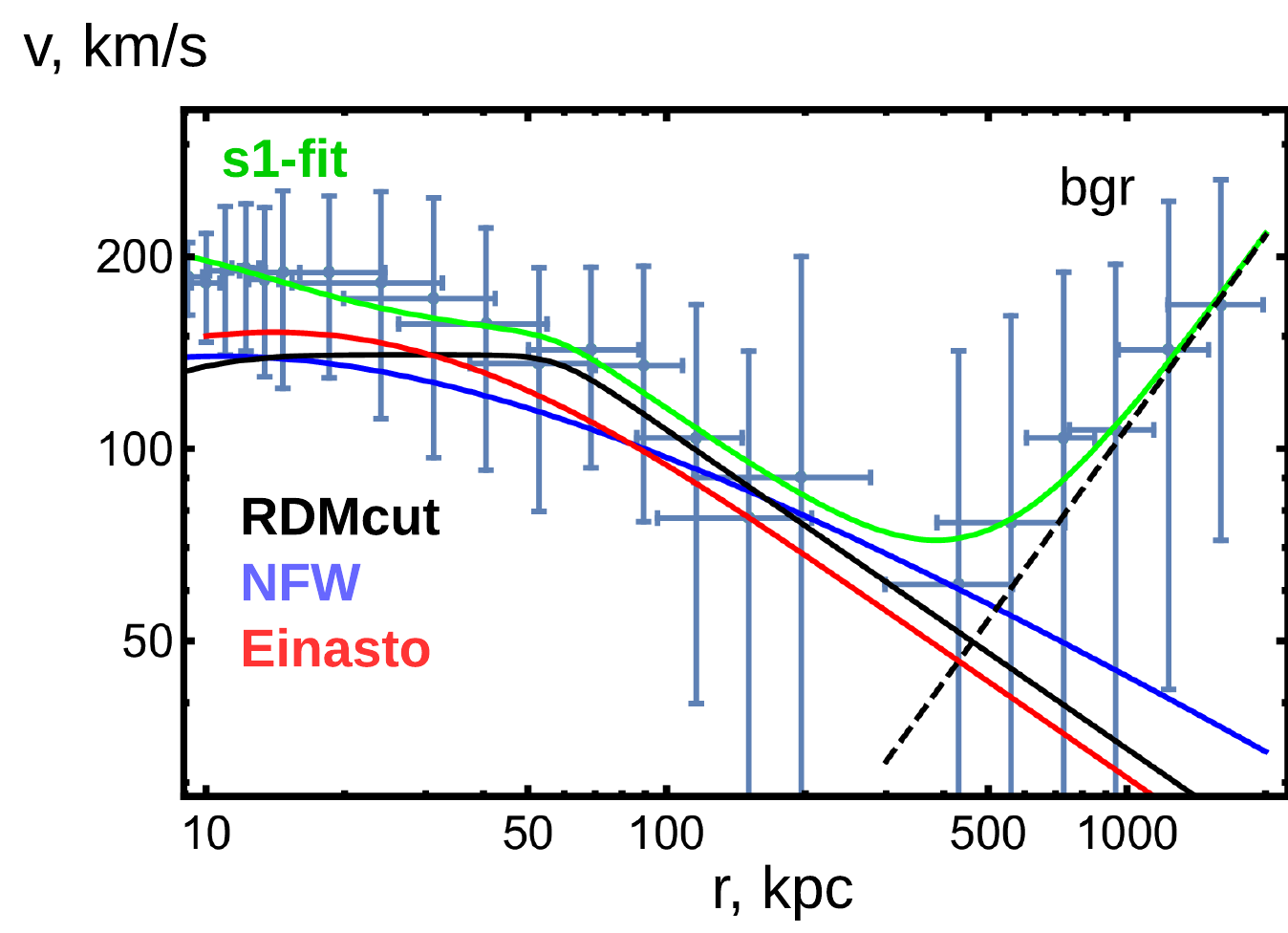}
\end{center}
\caption{The outer part of the rotation curve with different dark matter profiles.}\label{f5}
\end{figure}

On Fig.\ref{f5} we display these two profiles together with the smoothed cutoff of RDM density used in our model. The profiles are given by the formulae
\begin{eqnarray}
&&v^2_{NFW}\sim F_{NFW}(r/r_x)/r,\ F_{NFW}(x)=\log(1+x)-x/(1+x),\\ 
&&v^2_{E}\sim \gamma(3n,d_n(r/r_E)^{1/n})/r,\ d_n=3n-1/3+0.0079/n,
\end{eqnarray}
where $\gamma(n,x)$ is the lower incomplete gamma function. The parameters for Fig.\ref{f5} are selected to $r_x=5$; $n=3$, $r_E=21$. We see that the profiles closely follow each other in the corridor of the large data scatter.

Some remarks on the cutoff used in our model. We suppose that a transition from the radial to uniform type of dark matter distribution happens on the outer radius. This phenomenon can be similar to {\it a termination shock} in the transitional layer on the border of the solar system, where the radially directed solar wind meets the uniform interstellar medium. In this paper we do not model this transition exactly, but simulate it phenomenologically. For the exact modeling one needs to consider different types of radial dark matter, introduced in \cite{1701.01569}, including massive, null and tachyonic (M/N/T-RDM). We consider the case when the rays of dark matter are not strongly curved by the galactic gravitational field, so that RDM-stars appear as spherically symmetric radially diverging dark matter distributions. This definitely happens for N/T-RDM and those MRDM configurations, which correspond to the high velocity at infinity.

Technically speaking, these all are the types of hot dark matter, however it is a special case of {\it hot radial} dark matter, different from the usually considered {\it hot isotropic} dark matter by the absence of transverse components of pressure. It has been shown in \cite{1811.03368}, that the radial dark matter types, both the cold and the hot ones, explain the experimental rotation curves equally well. For the dark matter with isotropic pressure only the {\it cold isotropic} type explains the experiment, while the hot isotropic type produces non-physical highly relativistic rotation curves. This means that the argument by Barranco et al. in \cite{1301.6785} about the observed non-relativism of the experimental rotation curves rules out the hot isotropic dark matter, but does not exclude the hot radial one.

On the other hand, for the modeling of the termination shock one needs dark matter, stopping at a large distance, that is only possible for the {\it cold radial} type. Paper \cite{1701.01569} shows a threshold effect, the increase of density at the places where the dark matter stops and its trajectories have turning points. This is the prototype of the termination shock. The exact modeling, however, should take into account the bending of trajectories in the galactic gravitational field, which makes the problem much more complicated. This is the problem of finding generic massive geodesics in an axially symmetric self-consistent gravitational field. The approach taken in this paper is to consider spherically-symmetric RDM-stars with a phenomenologically imposed cutoff at a large distance, as {\it an approximation} to this generic problem.

\section{The disk contribution out of plane}

\begin{figure}
\begin{center}
\includegraphics[width=\textwidth]{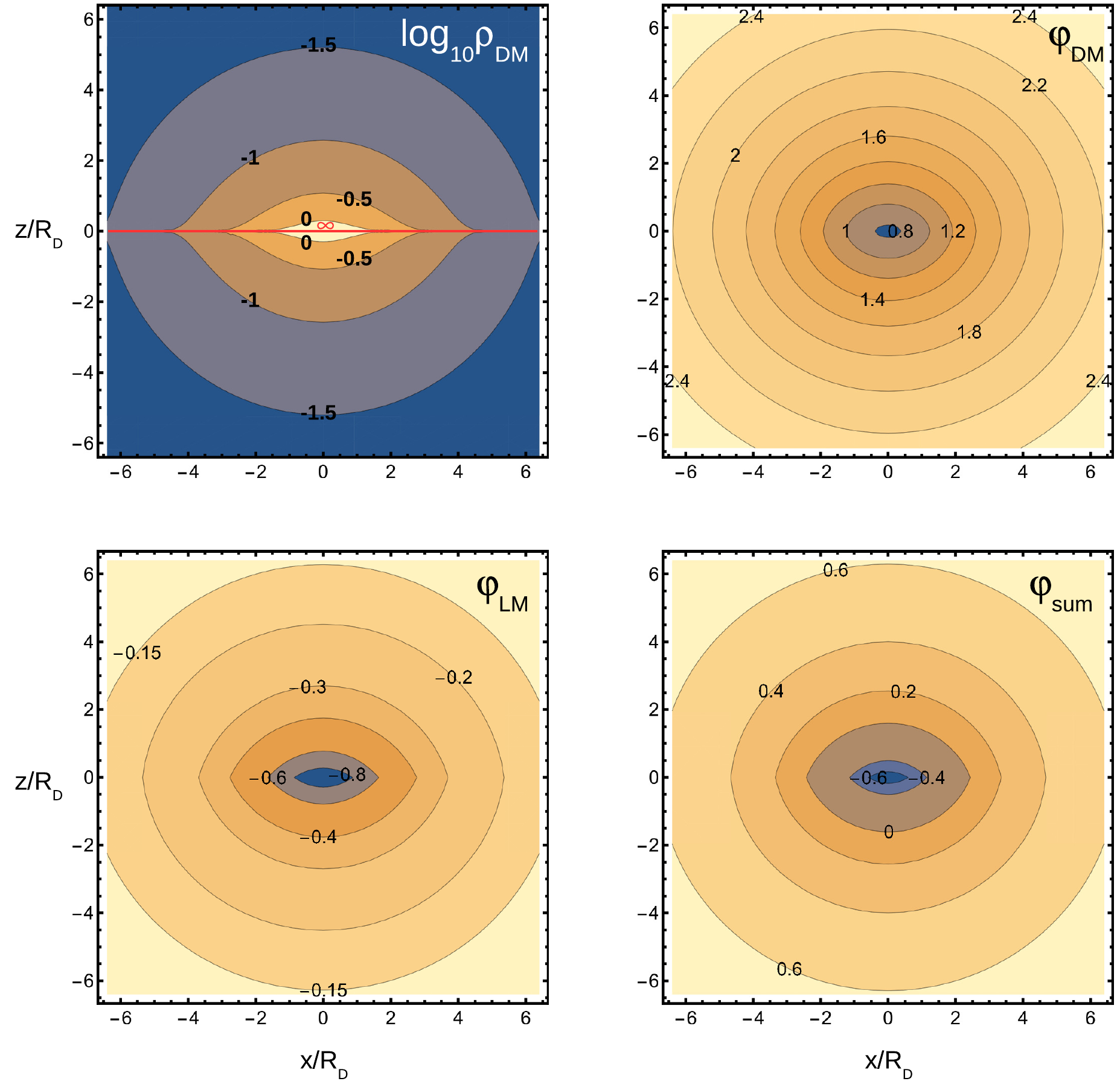}
\end{center}
\caption{The dark matter density and the disk contributions of dark and luminous matter to gravitational potential out of the galactic plane.}\label{f6}
\end{figure}

Let us compute the contribution of luminous mass disk out of a galactic plane:
\begin{eqnarray}
&&x=(r_1,0,z),\ y=(r\cos\alpha,r\sin\alpha,0),\\
&&\varphi_{dm}(x)=G/L_{KT} \int dM_{lm}(y)\,\log|x-y|\\
&&=G/L_{KT} \int_0^\infty r dr\, \Sigma_{lm}(r)\, I_4(r_1,z,r),\\
&&I_4(r_1,z,r)=\int_0^{2\pi}d\alpha/2\cdot\log(r^2+r_1^2-2rr_1\cos\alpha+z^2)\\
&&=\pi \log(1/2(r^2+r_1^2+z^2)(1+(1-4r^2r_1^2/(r^2+r_1^2+z^2)^2)^{1/2})),\\
&&\Sigma(r)=M_{lm}/(2\pi R_D^2)\exp(-r/R_D),\\
&&\varphi_{dm}(x)=GM_{lm}/(2\pi R_D^2 L_{KT}) \int_0^\infty r dr \exp(-r/R_D)I_4(r_1,z,r), \\
&&r_x=r_1/R_D,\ z_x=z/R_D,\ u=r/R_D,\ \tilde I_4(r_x,z_x,u)=\pi\log((u^2\\
&&+r_x^2+z_x^2)(1+(1-4u^2r_x^2/(u^2+r_x^2+z_x^2)^2)^{1/2}))\,(+Const),\\
&&\varphi_{dm}(r_x,z_x)=GM_{lm}/L_{KT}\int_0^\infty u du\, e^{-u}\tilde I_4(r_x,z_x,u)/(2\pi),\\
&&\varphi_{lm}(x)=-G\int dM_{lm}(y)/|x-y|,\\
&&=-G\int_0^\infty r dr\, \Sigma_{lm}(r)\, I_5(r_1,z,r)\\
&&=-GM_{lm}/(2\pi R_D^2)\int_0^\infty r dr \exp(-r/R_D)I_5(r_1,z,r),\\
&&I_5(r_1,z,r)=\int_0^{2\pi}d\alpha(r^2+r_1^2-2rr_1\cos\alpha+z^2)^{-1/2}\\
&&=2(K_e(-4rr_1/((r-r_1)^2+z^2))((r-r_1)^2+z^2)^{-1/2}\\
&&+K_e(4rr_1/((r+r_1)^2+z^2))((r+r_1)^2+z^2)^{-1/2}),\\
&&\varphi_{lm}(r_x,z_x)=-GM_{lm}/R_D \int_0^\infty u du\, e^{-u}I_5(r_x,z_x,u)/(2\pi),\\
&&\rho_{dm}(x)=\int dM_{lm}(y)/(4\pi L_{KT})/(x-y)^2\\
&&=M_{lm}/(8\pi^2 R_D^2 L_{KT})\int_0^\infty r dr \exp(-r/R_D)I_6(r_1,z,r),\\
&&I_6(r_1,z,r)=\int_0^{2\pi}d\alpha(r^2+r_1^2-2rr_1\cos\alpha+z^2)^{-1}\\
&&=2\pi(r_1^4 + 2 r_1^2 (-r^2 + z^2) + (r^2 + z^2)^2)^{-1/2},\\
&&\rho_{dm}(r_x,z_x)=M_{lm}/(4\pi R_D^2 L_{KT})\int_0^\infty u du\, e^{-u}I_6(r_x,z_x,u)/(2\pi),
\end{eqnarray}
where $K_e(x)$ is the complete elliptic integral of the first kind. We evaluate the internal integrals $I_{4-6}$ analytically and the external ones numerically. The resulting distributions are shown on Fig.\ref{f6}. All images are given in normalized coordinates $ (x / R_D, z / R_D) $. In the first image, the density is given in logarithmic scale, while in $ \rho_ {dm} $ the common dimensional factor $ M_ {lm} / (4 \pi R_D ^ 2 L_ {KT}) $ and in $ \log_ {10 } \rho_ {dm} $ the corresponding additive term are omitted. The second image shows the contribution to the gravitational potential $ \varphi_ {dm} $, with the omitted dimensional factor $ GM_ {lm} / L_ {KT} $. The third image is the contribution of $ \varphi_ {lm} $, where the factor $ GM_ {lm} / R_D $ is omitted. The fourth image shows the sum $ \varphi_ {sum} = \varphi_ {lm} + \varphi_ {dm} $, the common factor $ GM_ {lm} / R_D $ is omitted, and $ L_ {KT} / R_D = 3 $ is selected. All shown potentials are defined up to an additive constant.

Note the following property of $ \rho_ {dm} $. Centering the polar coordinate system around the point $ x = (r_1,0,0) $ and holding the leading term in the integral:
\begin{eqnarray}
&&\rho_{dm}(x)\sim\int_{+0} 2\pi rdr (\Sigma_{lm}(x)+...)/(4\pi r^2 L_{KT})\\
&&\sim \Sigma_{lm}(x)/(2L_{KT})\cdot\int_{+0} dr/r \to+\infty,
\end{eqnarray}
we see that the integral diverges logarithmically in the galactic plane. This feature is a memory of the original $ \delta (z) $ singularity of an infinitely thin disk. The singularity is cut off taking into account the finite thickness of the galactic disk. Also, the singularity would be erased when using the oscillating term introduced in \cite {0604496}:
\begin{eqnarray}
&&\rho_{dm}(x)\sim \int d^3x'\, \rho_{lm}(x')/|x-x'|^2\cdot(1-\cos(\mu|x-x'|)).
\end{eqnarray}
On the other hand, the contribution to the potential $ \varphi_ {dm} $ contains additional integration, which eliminates this singularity. Because of this, the dark matter potential $ \varphi_ {dm} $ in Fig.\ref {f6} is smooth. The potential of the luminous matter $ \varphi_ {lm} $ has a typical derivative discontinuity associated with a jump of the gravitational field when passing through the galactic plane in orthogonal direction. This jump is also an idealization eliminated when considering a galaxy of finite thickness. The sum of potentials inherits the features of both distributions. At large distances, all distributions become spherically symmetric, with the dominating contribution of dark matter.

\section{Conclusion}

We have studied the behavior of galactic rotation curves in the model of RDM-stars. First of all, we considered the contribution to the gravitational field of dark matter, calculated using KT-integral for a corpuscle of luminous matter, or, equivalently, the contribution of dark matter from a single RDM-star. The resulting gravitational field gives a flat rotation curve. Further, taking into account the distribution of luminous matter in KT-integral or the distribution of RDM-stars in the galaxy, non-planar curves are obtained. Taking the distribution of RDM-stars, proportional to the density of luminous matter, allows to obtain a good fit of experimental data, both for the universal rotation curve describing general spiral galaxies and for the grand rotation curve describing the Milky Way in a wide range of distances.

\paragraph*{Acknowledgments.} Many thanks to Paolo Salucci for his valuable remark on the behavior of experimental rotation curves, which has inspired the author for writing this paper. The author also thanks Kira Konich for proofreading the paper.

\end{document}